\PassOptionsToPackage{dvipsnames}{xcolor}
\documentclass[preprint]{aastex631}

\usepackage{amsmath}
\usepackage{physics}
\usepackage{cleveref}
\usepackage{bm}
\usepackage{capt-of}   
\usepackage{svg}
\usepackage{empheq}
\usepackage{enumitem}
\usepackage{float}
\usepackage{hyperref}
\usepackage{hhline}
\usepackage{longtable}
\usepackage{numprint}
\usepackage{pifont}
\usepackage{xspace}
\usepackage{placeins}
\usepackage{graphicx}
\usepackage{mwe}
\usepackage{multirow}
\usepackage{tikz}
\usepackage{listings}

\usetikzlibrary{positioning, arrows.meta}

\newcommand{\ho}{\ensuremath{H_0}\xspace}

\newcommand{\kl}{\normalfont \text{KL}\xspace}

\newcommand{\gigal}{\texttt{GIGA-Lens}\xspace}
\newcommand{\gigalone}{\texttt{GIGA-Lens 1.0}\xspace}
\newcommand{\gigaltwo}{\texttt{GIGA-Lens 2.0}\xspace}
\newcommand{\inspec}{\texttt{in\_spec}\xspace}
\newcommand{\outspec}{\texttt{out\_spec}\xspace}
\newcommand{\shardmap}{\texttt{shard\_map}\xspace}
\newcommand{\twopr}{\ensuremath{^{\prime \prime}}}

\newcommand{\rhat}{\ensuremath{\hat{R}}\xspace}

\newcommand{\hst}{\emph{HST}\xspace}

\newcommand{\ser}{S{\'e}rsic\xspace}

\newcommand{\desitwothreeeight}{DESI~J238.5690+04.7276\xspace}

\submitjournal{ApJ}

\shorttitle{\gigaltwo}
\shortauthors{Huang, Upson, Ratier-Werbin, Lu, Yap, Xu, Odell, Parke et al.}

\begin{document}
\title{\gigaltwo: Strong-Lens Modeling on Multiple GPU \emph{Nodes}
}

\correspondingauthor{
Linus Upson, Ansel Parke, Xiaosheng~Huang}
\email{linusu@berkeley.edu, aparke@berkeley.edu,
xhuang22@usfca.edu}

\author[0000-0001-8156-0330]{Xiaosheng~Huang}
\affiliation{Department of Physics \& Astronomy, University of San Francisco, San Francisco, CA 94117, USA}
\affiliation{Physics Division, Lawrence Berkeley National Laboratory, 1 Cyclotron Road, Berkeley, CA 94720, USA}

\author[0009-0003-6028-5677]{Linus Upson}
\affiliation{Department of Physics, 
University of California, Berkeley, CA 94720, USA}

\author[0009-0009-8206-0325]{Nicolas~Ratier-Werbin}
\affiliation{Department of Physics, 
Complutense University of Madrid, 28040 Madrid, Spain}
\affiliation{Physics Division, Lawrence Berkeley National Laboratory, 1 Cyclotron Road, Berkeley, CA 94720, USA}

\author{Harry Lu}
\affiliation{Department of Electrical Engineering and Computer Science, 
University of California, Berkeley, CA 94720, USA}

\author{Sean~Xu}
\affiliation{Department of Physics, University of California, Berkeley, CA 94720, USA}
\affiliation{Physics Division, Lawrence Berkeley National Laboratory, 1 Cyclotron Road, Berkeley, CA 94720, USA}

\author[0009-0009-3646-8944]{Elden Yap}
\affiliation{Department of Electrical Engineering and Computer Science, 
University of California, Berkeley, CA 94720, USA}

\author[0009-0006-9722-2401]{Evan Odell}
\affiliation{Department of Astronomy, University of California, Berkeley, CA 94720, USA}
\affiliation{Physics Division, Lawrence Berkeley National Laboratory, 1 Cyclotron Road, Berkeley, CA 94720, USA}

\author{Ansel Parke}
\affiliation{Department of Physics, 
University of California, Berkeley, CA 94720, USA}


\author[0009-0003-5592-3515]{Harsh Ambardekar}
\affiliation{Department of Physics, University of California, Berkeley, CA 94720, USA}
\affiliation{Physics Division, Lawrence Berkeley National Laboratory, 1 Cyclotron Road, Berkeley, CA 94720, USA}

\author[0009-0003-4697-7079]{Saul~Baltasar}
\affiliation{Department of Physics, 
Complutense University of Madrid, 28040 Madrid, Spain}
\affiliation{Physics Division, Lawrence Berkeley National Laboratory, 1 Cyclotron Road, Berkeley, CA 94720, USA}

\author[0000-0002-0530-6530]{Nestor~Demeure}
\affiliation{National Energy Research Scientific Computing Center,
Berkeley, CA 94720, USA}

\author{Bradley~Richardson}
\affiliation{National Energy Research Scientific Computing Center,
Berkeley, CA 94720, USA}

\author[0000-0003-2748-7333]{Andi~Gu}
\affiliation{Quantum Science and Engineering, Harvard University, 
Cambridge, MA 02138, USA}

\author[0000-0002-6876-8492]{Yuan-Ming~Hsu}
\affiliation{Department of Physics, National Taiwan University, Taipei City 106319, Taiwan}

\author{Junyi~Liu}
\affiliation{Department of Physics, Rose-Hulman Institute of Technology, Terre Haute, IN 47803, USA}

\begin{abstract}

We present \gigaltwo: a major upgrade to the GPU-accelerated Bayesian framework for modeling strong lensing systems that allows it to be run across multiple GPU \emph{nodes}. 
We have succeeded in running \gigaltwo on 128 nodes or 512 A100 GPUs. We demonstrate the speed benefits of this new version, and apply them to modeling 100 simulated systems and a real system, \desitwothreeeight. We also present other changes to the framework that have yielded further improvement on performance.
\end{abstract}

\section{Introduction}\label{sec:intro}
Strong gravitational lensing has emerged as a powerful and versatile probe of cosmology, providing direct sensitivity to the nature of dark matter and dark energy, and the Hubble constant (\ho). Small-scale perturbations in lens mass distributions encode information about the dark matter particle mass spectrum and subhalo population \citep[e.g.,][]{dalal2002, vegetti2010a}, while time-delay measurements in lensed quasars and supernovae offer an independent and competitive route to measuring $H_0$ \citep[e.g.,][]{refsdal1964a, treu2016a}. 
Strong lensing systems with multiple source planes can constrain cosmological models \citep[e.g.,][]{jullo2010a, collett2014a, caminha2022a, urcelay2026a} through their sensitivity to ratios of angular-diameter distances.

Realizing this scientific potential requires fully Bayesian inference with demonstrably reliable posterior sampling. For all three science goals---detection of low-mass dark matter (sub)halos, time-delay cosmography, and dark energy constraints---restricting the lens-model parameter space or failing to marginalize over key sources of uncertainty can propagate directly into biased or overly confident conclusions, or both \citep[e.g.,][]{Suyu2010a,Birrer2020a}. As lens models become more flexible and complex and lens samples grow in size, ensuring robust posterior exploration and numerical convergence becomes critically important, placing stringent demands on computational performance, parallelism, and scalability.

The \gigal \citep[][G22]{gu2022a} framework was developed to address this challenge by combining GPU-accelerated full forward modeling with explicit posterior sampling convergence diagnostics, introducing quantitative convergence metrics such as the Gelman-Rubin metric, $\hat R$, and effective sample size to strong lens modeling.
This approach enabled statistically robust modeling of strong lensing systems, but its execution in its original implementation was limited to single-node GPU configurations, hindering its scalability to large sample sizes or large and complex lensing systems.
The original \gigal from G22 will be henceforth referred to as \gigal~\texttt{1.0}.

In this work, we introduce \gigaltwo, a major upgrade to the \gigal framework that extends fully forward Bayesian strong-lens inference across multiple GPU nodes. 
We achieve unprecedented scaling up to 128 nodes, or 512 A100 GPUs---two orders of magnitude more than for \gigalone---and quantify the performance gains.
As a demonstration, we apply \gigaltwo to both large ensembles of simulated systems and a representative observed strong lens. For the next generation of wide-field imaging surveys, the limiting factor for cosmological inference from strong lensing will not be data volume, but our ability to model lenses robustly and at scale. \gigaltwo overcomes this limitation by enabling high-fidelity Bayesian inference at unprecedented scale without sacrificing statistical rigor.

Furthermore, gaining speed without sacrificing statistical rigor allows us to model the same system multiple times under different modeling assumptions to determine systematic uncertainties.

This paper is organized as follows. In \S\,\ref{sec:need}, we describe the motivation. 
\S\,\ref{sec:methodology} presents the multi-node parallelization strategies implemented in \gigaltwo, including distributed computation for each of the three modeling stages: multi-start MAP estimation, stochastic variational inference (SVI), and Hamiltonian Monte Carlo (HMC) sampling, along with the associated convergence diagnostics. 
In \S\,\ref{sec:results}, we demonstrate the performance of our pipeline by modeling 100 simulated systems. 
We also show the results of applying \gigaltwo to a real strong lens with \hst imaging shown in Appendix~\ref{sec:real-sys}. 
We conclude in \S\,\ref{sec:conclusion}. 

\FloatBarrier
\section{Motivation}\label{sec:need}
Modern strong lensing inference requires fully forward Bayesian modeling
with rigorous uncertainty quantification,
but achieving this level of robustness
imposes substantial computational demands.
By moving to a multi-node execution model,
we gain access to both substantially increased computational throughput
and a greatly expanded memory footprint.
Together, these capabilities are essential
for robust and scalable strong lensing inference,
and they benefit all stages of the \gigal modeling pipeline,
as discussed in detail in the remainder of this paper.

Lens modeling is a highly nonlinear problem,
and although a single forward simulation---including backpropagation
for gradient evaluation---can be executed rapidly on modern GPUs,
adequately exploring the posterior requires drawing a large number of samples
in a high-dimensional parameter space.
Both the computational cost and memory requirements
grow rapidly with increasing image resolution,
the Einstein radius of the system,
and overall model complexity,
encompassing the structure of the source light,
the main lens mass model
(which, for group- or cluster-scale lenses, typically
includes at least some of the member galaxies),
and the inclusion of field galaxies
in light and, where relevant, mass.
For multi-lens-plane and/or multi-source-plane systems,
the complexity increases further.
As a result, for accurate inference in realistic systems, 
achieving converged posterior sampling
often becomes the dominant computational bottleneck.

For cosmological inference, beyond statistical rigor,
speed is also critical for assessing and controlling
systematic uncertainties.
Testing the impact of different model assumptions
requires performing the full modeling procedure
for the same system multiple times,
often with alternative parameterizations,
priors, or data treatments.
Accelerating inference therefore enables
systematic exploration of modeling choices.

This capability is especially important for large,
complex systems that are also particularly valuable
for cosmology, such as the Carousel Lens, which is a cluster-scale lensing system \citep{sheu2024a, odonnell2026a, urcelay2026a} and demands orders of magnitude more computation
than galaxy-scale lenses
when modeled using a fully forward Bayesian approach.
For comparison, for modeling galaxy-scale strong lenses, the typical cutout size for \hst observations is $\mathcal{O}(10^4)$ pixels \citep{huang2026a, huang2025c} vs. $\mathcal{O}(10^6)$ pixels for the Carousel Lens, and in terms of parameters, several tens for the former and hundreds for the latter.
The same considerations apply to large samples
of galaxy-scale strong lenses (e.g., in a hierarchical Bayesian framework),
where modeling each system with the full complexity it requires,
rather than relying on simplifying assumptions (e.g., SIE instead of EPL, fixing the \ser index instead of fitting for it).
The subsequent assessment of systematics across the entire sample
becomes even more computationally prohibitive.
Finally, joint analyses that combine lensing
with additional information such as stellar dynamics
dramatically increase both computational
and memory requirements.
In all of these cases,
greater computational capacity translates directly into more thorough posterior exploration
and more reliable physical conclusions.

In this work, we address these challenges
by further accelerating \gigal on multiple GPU nodes
of the \emph{Perlmutter} supercomputer
at the National Energy Research Scientific Computing Center (NERSC).\footnote{\url{https://www.nersc.gov/}}
By exploiting multi-node parallelism,
we enable inference workflows
that were previously infeasible
within practical time or memory constraints
on a single GPU node.

Our implementation leverages JAX’s \shardmap\ API
to distribute computation across devices (i.e., individual GPUs) and nodes
while maintaining a unified programming model.
In addition to algorithmic advances,
we have substantially improved the documentation
and execution environment,
providing clear examples
and reproducible setup instructions.
Together, these developments make scalable,
rigorous strong lens modeling
both powerful and accessible.

\section{Methodology}\label{sec:methodology}
\gigal is a fully forward-modeling Bayesian lens modeling pipeline.
Briefly, it consists of three stages: finding the maximum a posteriori (MAP) for the lensing parameters via multi-start gradient descent, determining a surrogate multidimensional Gaussian covariance matrix for these parameters using stochastic variational inference (SVI), and finally sampling with Hamiltonian Monte Carlo (HMC).
All three stages use gradient descent with automatic differentiation and take advantage of GPU acceleration.  

Our mass model consists of an elliptical power law (EPL) for the lens mass profile and external shear. 
The Einstein radius is denoted $\theta_E$ (arcsec), and $\gamma$ is the EPL mass-slope parameter.
$(x, y)$ denote the center coordinates, specified independently for the lens mass, lens light, and source light. 
($\gamma_{1, \text{ext}}$, $\gamma_{2, \text{ext}}$) are the external shear components.
The parameters $\epsilon_1$ and $\epsilon_2$ specify the eccentricities.
We model lens light with one or more \ser profiles.
For source light, we use \ser profiles and/or shapelets \citep{birrer2015a}.
For the lens and source surface brightness scaling, we either model the light intensities as parameters (``forward'' modeling) or solve for them by linear inversion (``backward'' modeling).
Both are mathematically justified.

We present our parallelization implementation in \S\,\ref{sec:parallel}.
We then present two improvements on the software side: a better optimizer choice for SVI in \S\,\ref{sec:optimzer} and mass matrix adaption and in \S\,\ref{sec:mass-matrix}. 
Finally we discuss sampling convergence in \S\,\ref{sec:convergence}.

\subsection{Parallelization Strategies}\label{sec:parallel}
Parallelization for the MAP and HMC stages are trivial, which will be presented first. We then go into the necessary detail for SVI. Throughout this work, a device refers to a single GPU, i.e., they are synonymous. 

\emph{Parallelization for MAP and HMC} \: Because each particle in the MAP performs gradient descent independently, parallelization across multiple GPUs is as simple as running a fraction of the particles on each GPU.\footnote{Previously, we ran a gradient descent on the mean of all particles' losses. By the chain rule, this is mathematically equivalent.}
We parallelize HMC in a similar way. Each GPU runs an independent set of chains, which are then aggregated once the sampling has completed.

\emph {Parallelization for SVI}  \: Stochastic Variational Inference (SVI) is more challenging to parallelize
than either MAP or HMC,
because it requires coordinated gradient updates
across devices.
In contrast to MAP particles or HMC chains,
which evolve independently,
SVI optimizes a \emph{shared} set of surrogate posterior parameters.
Updating these parameters requires an averaging operation across all devices,
which in turn necessitates communication among GPUs,
including GPUs residing on different compute nodes.

We follow the same SVI formulation as in G22,
which we briefly summarize below.
SVI fits a multivariate normal surrogate posterior
$\tilde{q}(\tilde{\Theta}; \tilde{\mu}, \tilde{\Sigma})$
to the true posterior
$\tilde{p}(\tilde{\Theta} \mid \mathcal{I}_{obs})$.
Here, $\mathcal{I}_{obs}$ is the observed image.
The parameter vector $\Theta$ refers to the full set of model parameters, while $\tilde{\Theta}$ denotes the corresponding parameters mapped into an unconstrained space via a bijective transformation (see G22 for details).
Throughout, the $\tilde{.}$ notation indicates quantities defined in the unconstrained space. In particular, $\tilde{p}$, $\tilde{q}$, and the surrogate mean $\tilde{\mu}$ and covariance $\tilde{\Sigma}$ all refer to distributions or parameters expressed in this transformed space.
 
We find the optimal surrogate 
by minimizing the Kullback--Leibler (KL) divergence
between the two distributions:

\begin{equation}
    \tilde{\mu}^*_{VI}, \tilde{\Sigma}^*_{VI} = \underset{\tilde{\mu}, \tilde{\Sigma}}{\text{argmin}} \ \kl (\tilde{q}(\tilde{\Theta}; \tilde{\mu}, \tilde{\Sigma}) \mid \mid \tilde{p}(\tilde{\Theta} \mid \mathcal{I}_{obs})) \label{eqn:vi}
\end{equation}

\noindent
where $\tilde{\mu}^*_{VI}$ and $\tilde{\Sigma}^*_{VI}$ are the mean and covariance matrix of the optimal surrogate.

The KL divergence is evaluated by expressing it
as an expectation over the surrogate posterior.
This formulation yields an objective function, the Evidence Lower Bound (ELBO) loss,

\begin{equation} \label{eqn:elbo}
        \mathrm{ELBO} = \mathbb{E}_{\tilde{\Theta}}[\log \tilde{q}(\tilde{\Theta}; \tilde{\mu}, \tilde{\Sigma})-\log \tilde{p}(\mathcal{I}_{obs},\tilde{\Theta})],
\end{equation}

\noindent
where $\tilde{\Theta}$ is sampled from the surrogate $\tilde{q}(\tilde{\Theta}; \tilde{\mu}, \tilde{\Sigma})$. In practice, this expectation value is approximated by averaging over a finite number of samples, $n_{VI}$, at each iteration.

The gradient of the ELBO with respect to the surrogate parameters ($\tilde{\mu}, \tilde{\Sigma}$) can also be expressed as an expectation value, 

\begin{equation} \label{eqn:elbo-grad}
    \grad_{\tilde{\mu}, \tilde{\Sigma}} \text{ELBO} = \mathbb{E}_{\tilde{\Theta}}\qty[\qty(\log \tilde{q}(\tilde{\Theta}; \tilde{\mu}, \tilde{\Sigma})-\log \tilde{p}(\mathcal{I}_{obs}, \tilde{\Theta})) \grad_{\tilde{\mu},\tilde{\Sigma}} \log \tilde{q}(\tilde{\Theta}; \tilde{\mu}, \tilde{\Sigma})]
\end{equation}

\begin{figure}
    \centering
    \includegraphics[width=0.95\textwidth] {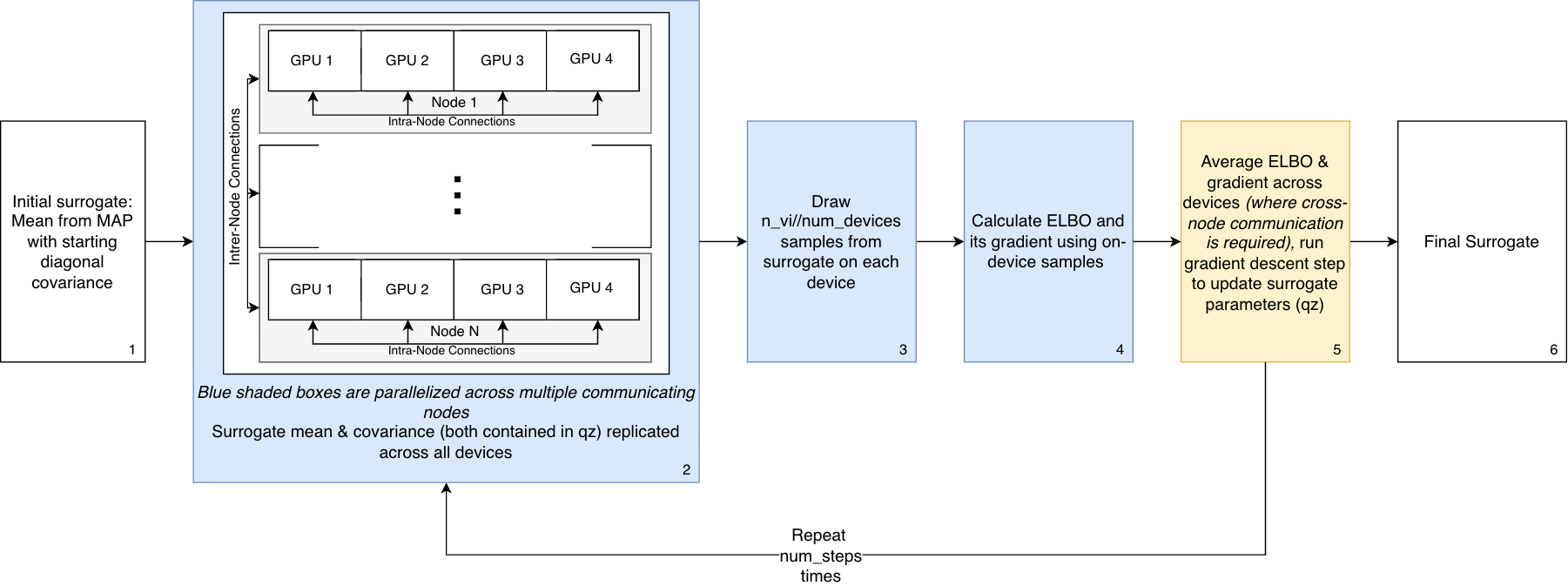} 
    \caption{Schematic of distributed computation for the SVI stage of the \gigaltwo pipeline, broken down into six steps, each represented by a box.
    We show $N$ GPU nodes with a total of 4$N$ GPUs. 
    A device is synonymous with a single GPU.    
    Each step is represented by a box; the steps where computation takes place on individual devices are shown as blue and the yellow box is the step where communications among all devices, including across nodes, are required. The communication among the GPUs on one node is accomplished via JAX \texttt{pmap}, whereas the communication across multiple GPU \emph{nodes} is accomplished via JAX \shardmap. 
    All the GPUs are shown in box~2. This box is replicated in steps 3 and 4, but for brevity, we do not show individual GPUs. The communication among devices and across nodes takes place in step 5, the yellow box. So far we have successfully run \gigaltwo on a total of $N=128$ nodes, or 512~GPUs.}
    \label{fig:ParalellizationFlowcharts}
\end{figure}

Computationally, we approximate $\grad_{\tilde{\mu}, \tilde{\Sigma}} \text{ELBO}$ by drawing $\left \lfloor{\frac{n_{VI}}{\mathrm{num\_devices}}}\right \rfloor $ samples from $\tilde{q}(\tilde{\Theta}; \tilde{\mu}, \tilde{\Sigma})$ on each device. With JAX's automatic differentiation, we separately estimate $\grad_{\tilde{\mu}, \tilde{\Sigma}} \text{ELBO}$ using each device's samples. The average of these gradients across all devices is an estimate of $\grad_{\tilde{\mu}, \tilde{\Sigma}} \text{ELBO}$ from a total of $n_{VI}$ samples. We can then minimize the ELBO loss using an gradient descent-based optimizer of our choice.\footnote{This approximation using a finite sample size yields a stochastic estimate of the ELBO gradient. Optimization with such gradient estimates is commonly referred to as stochastic variational inference \citep{hoffman2013}, or SVI.} This is the most challenging part of the multi-node implementation of this pipeline due to cross-node communication (see Fig \ref{fig:ParalellizationFlowcharts}).

In each iteration, most of the computational cost comes from simulating the samples drawn from the posterior in the computation for ELBO its gradients. 
We draw an equal fraction of the total samples on each device and then run the full calculation of ELBO and its gradient without communication between devices. Gradients are then aggregated across devices and an optimizer step is applied to the parameters ($\tilde{\mu}, \tilde{\Sigma}$) of the Gaussian surrogate. At the end of each iteration, the parameters are updated and replicated across devices, so that they remain identical across devices.

\subsection{Choice of Optimizer for SVI}\label{sec:optimzer}
\begin{figure}[H]
    \centering
    \includegraphics[width=0.4\textwidth]{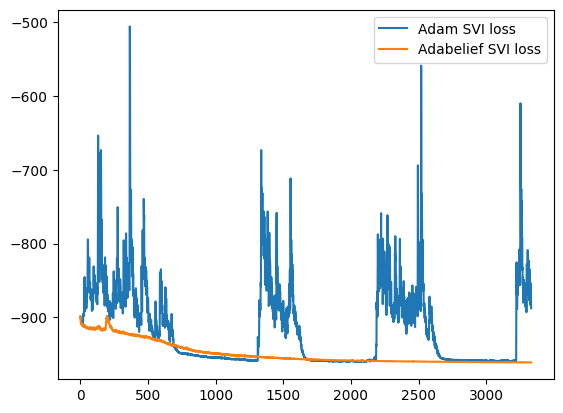}
    \caption{Sample loss curves for Adam compared to AdaBelief. AdaBelief is able to achieve a lower loss in a quicker and stabler manner. In our experience, sampling longer for AdaBelief typically leads to even better surrogates, while the same is not true for Adam.}
    \label{fig:AdaAdamLoss}
\end{figure}

\begin{figure}[H]
    \centering
    \includegraphics[width=1\textwidth]{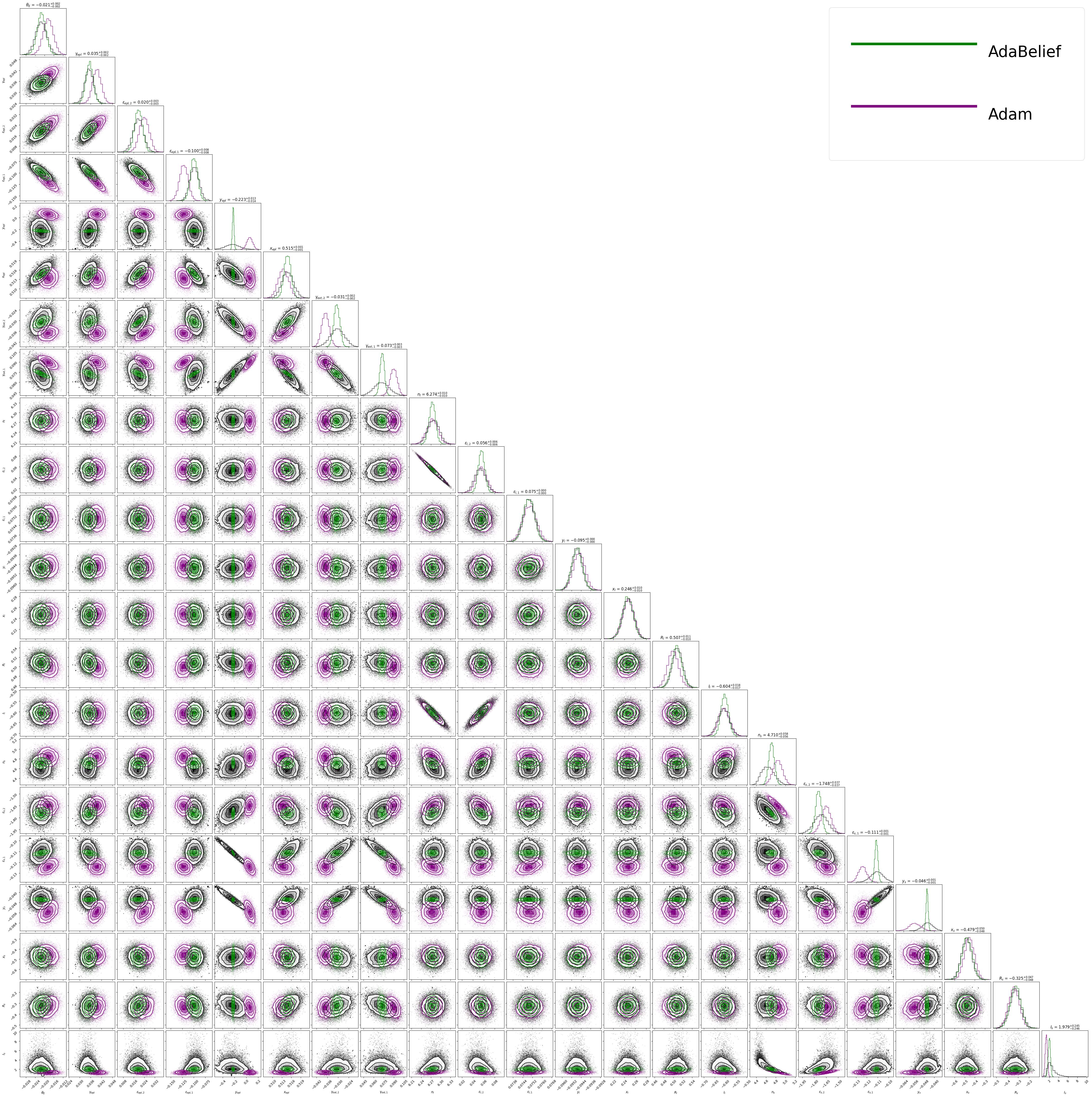}
    \caption{A representative corner plot for surrogates generated by Adam (purple) and AdaBelief (green) for SVI, compared with the posterior from HMC (black). Typically, AdaBelief's surrogate provides a better approximation to the true posterior for most parameters over Adam, both in terms of the mean and covariance. 
    }
    \label{fig:AdaAdamCorner}
\end{figure}

In modeling high resolution real systems \citep{huang2026a, huang2025c} and in the development of this version of the pipeline, we encountered a number of systems where the convergence for HMC was slow. 
We identified the reason to be the best-fit surrogate from SVI being far from a good approximation to the true posterior.
For SVI, as with MAP, we use the Adam optimizer as the default. 
We notice that that, with Adam,
the ELBO loss for SVI often sharply increases to much higher than its starting value for periods of a few hundred epochs, before once again returning to more reasonable values for a time. This cycle could sometimes repeat multiple times during SVI, 
suggesting that Adam was repeatedly taking the parameters out of a minimum in the loss landscape.
This was supported by our tests using a simpler optimizer, Stochastic Gradient Descent. With higher learning rates, SGD exhibited similar behavior, while with lower learning rates, it converged much more stably. Despite Adam's difficulty converging in certain cases, we still wanted the benefits of a more advanced optimizer than SGD. 
We therefore adopt AdaBelief \citep{zhuang2020}.
AdaBelief adapts its step size according to the deviation of the current gradient from its recent average, effectively scaling updates based on gradient predictability. By tracking gradient deviations, it reduces the impact of noisy or outlier gradients in regions of high variance. This adaptive behavior leads to more robust optimization in practice (Figure~\ref{fig:AdaAdamLoss}). 
In our experiments, AdaBelief consistently exhibits greater stability and provides provides a closer approximation to the true posterior than Adam (Figure~\ref{fig:AdaAdamCorner}).
AdaBelief now is the default optimizer for SVI with these hyperparameters: $\texttt{lr=1e-4, b1=0.95, b2=0.99}$.
We have not exhaustively explored AdaBelief's hyperparameter space, but our experience finds these values to be generally effective. 
We finally note that unlike Adam, we do not typically apply a learning rate schedule to AdaBelief.

For the MAP, whether we use Adam or AdaBelief does not appear to make a notable difference.

\subsection{Mass Matrix Adaptation in HMC}\label{sec:mass-matrix}

The purpose of the SVI stage is to properly initialize the sampling stage using HMC. 
The covariance matrix of the surrogate posterior from SVI provides the approximate scale for the true posterior in each dimension. 
For most lensing systems, this ensures a reasonable acceptance rate in HMC, which results in efficient exploration and less-correlated samples, yielding faster sampling convergence.
We have found, however, there are especially challenging systems, simulated and real, for which SVI, even with AdaBelief, provides a poor surrogate.
Therefore, in \gigaltwo, we implement a new mass matrix adaptation \citep[e.g.,][]{stan2021} step to provide a better estimate of the covariance matrix: we use larger-than-typical ($\sim 10 \times$) burn-in steps, to the last 90\% of which we fit a new Gaussian surrogate. This is then used to initialize HMC.
In this implementation, no additional burn-in is performed prior to final sampling. A second burn-in phase, in which the step size is retuned to the updated mass matrix, could potentially yield improved sampling performance. 
We leave this exploration for future work.

\subsection{Convergence Metrics}\label{sec:convergence}
Before demonstrating the performance of our pipeline on simulated systems in the next section, we introduce the convergence metrics used throughout this work.

We employ two widely used diagnostics: the potential scale reduction factor, the Gelman-Rubin metric, \rhat, and the effective sample size (ESS). The ESS provides an estimate of the number of effectively independent samples obtained from Hamiltonian Monte Carlo (HMC), with higher values indicating better sampling performance.

The suggested threshold for convergence is $\rhat < 1.1$ \citep{gelman2014a}.\footnote{A discrepancy exists between the current definition of $\hat{R}$ \citep{gelman2014a} and its implementation in TensorFlow Probability \citep{brooks1998a}: the former is effectively the square root of the latter; see \citet{huang2025c} for details.} 
We meet this criterion for all systems modeled in this work, including both simulated and real lenses.

For ESS, \citet{vehtari2021} emphasized the importance of accounting for cross-chain variance: sampling multiple chains is not equivalent to simply summing the ESS across chains, and neglecting this effect can lead to overestimation of the effective sample size.
This accounting is implemented in TensorFlow Probability by appropriately setting the \texttt{cross\_chain\_dims} argument (rather than using the default value of \texttt{None}). 
All reported ESS values are calculated using this definition, and in all cases we satisfy the recommended threshold.



\section{Results}\label{sec:results}

\subsection{Speed and Scaling}

\begin{figure}[H]
    \centering
    \includegraphics[width=0.75\linewidth]{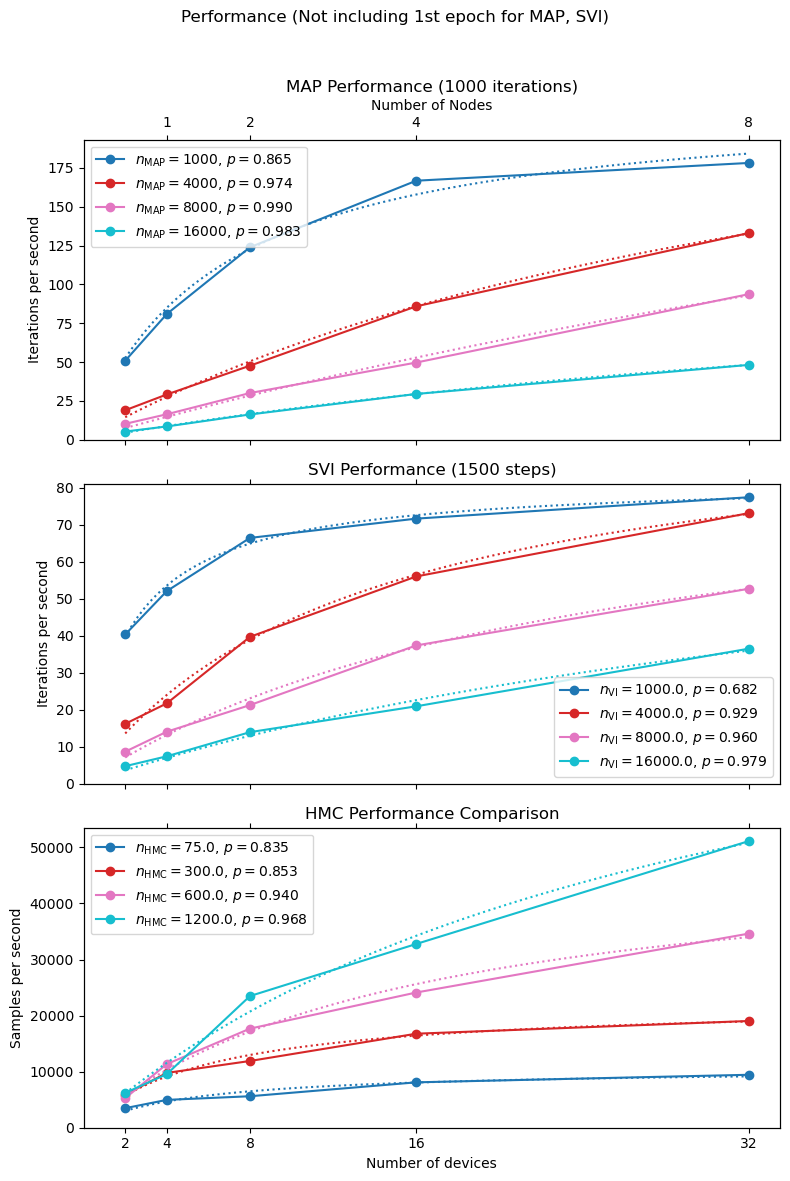}
    \caption{Speeds of the three pipeline stages vs. the number of GPU Nodes. Note that for HMC, burn-in phase is not included. Dotted lines are Amdhal's Law fits.}
    \label{fig:performance1ep}
\end{figure}

Parallelization across many compute nodes yields great improvements in the speed of the \gigal pipeline. To measure this improvement, we modeled a sample system on 2, 4, 8, 16, and 32 A100 GPUs with a range of pipeline hyperparameters (See Table \ref{tab:speedtestparams}, Figure~\ref{fig:performance1ep}). Because the first iteration of the inference algorithms includes JIT compiling functions, as well as other one-time computational costs (``overhead''), we do not include it in the timing plots displayed here. If it is included, it adds a constant $\sim 5$s for MAP, $\sim 10$ – 20~s for SVI and HMC). 
We treat the burn-in steps for HMC in the same way. Because burn-in can take a variable amount of time, we only count the time for the final sampling.

\begin{deluxetable}{ |c|c|c|c|c|c| }[H]
\tablecaption{Pipeline Hyperparameters for Multi-node Speed Measurements} \label{tab:speedtestparams}
\tablehead{
    \multicolumn{2}{c}{MAP} &
    \multicolumn{2}{c}{SVI} &
    \multicolumn{2}{c}{HMC}
}
\startdata
Steps & $n_\mathrm{MAP}$ & Steps & $n_\mathrm{SVI}$ & Steps &$n_\mathrm{HMC}$ \\ \hline
 \multirow{4}{4em}{1000} & 1000 & \multirow{4}{4em}{1500} & 1000 & \multirow{4}{4em}{750} & 75 \\
 & 4000 & & 4000 & & 300\\
 & 8000 & & 8000 & & 600\\
 & 16000 & & 16000 & & 1200\\
\enddata
\end{deluxetable}

We choose the hyperparameters for the pipeline to be representative of what we use for modeling real systems. For MAP, while 350 steps (used in G22) is sufficient for most simulated systems, some of them require more steps; thus we run it for 1000 to ensure the global minimum is reached. Furthermore, for real systems, we sometimes need even more steps \citep{huang2025c}. 
For SVI, although fewer iterations are often sufficient, we allow up to 1500 steps based on our experience with real systems. 
Finally, for HMC, in the current implementation, 1200 chains represent a very generous upper limit.\footnote{There is research indicating that using many short chains can yield advantages \citep[e.g.,][]{margossian2021}. However, this needs to be thoroughly investigated before potential incorporation into our pipeline, and is left for future work.}

We also include a fit of Amdahl’s Law \citep{amdahl1967} to the data (Figure~\ref{fig:performance1ep}). The fitted parallelizable fraction $p$ quantifies how efficiently additional GPUs translate into speedup ($p=1$ would mean perfect linear scaling, while lower values indicate diminishing returns from adding devices): for the largest workloads tested, $p$ reaches 0.983 for MAP ($n_{\mathrm{MAP}} = 16000$), 0.979 for SVI ($n_{\mathrm{VI}} = 16000$), and 0.968 for HMC ($n_{\mathrm{HMC}} = 1200$), confirming that all three pipeline stages exceed 96\% parallel efficiency when the per-device workload is sufficiently large. Even at more modest workloads (e.g., $n_{\mathrm{MAP}} = 4000$, $n_{\mathrm{VI}} = 4000$, $n_{\mathrm{HMC}} = 300$), $p$ remains above 0.85 for all three stages. As expected, $p$ increases with the number of simultaneous tasks per iteration, since a larger workload amortizes the fixed overhead of inter-node communication.

\clearpage
\subsection{Hundred Simulated Systems}\label{sec:100-sys}
The ability to parallelize across multiple GPU nodes allowed us to run inference much faster, and we can circumvent many of the memory or speed limitations that otherwise would arise. 
On the software side, we have also made improvements (see \S\,\ref{sec:optimzer} and \ref{sec:mass-matrix}).
As in G22, we test our new pipeline on 100 simulated systems on 16 GPU devices (four GPU nodes). 
We modeled them with an 
80-pixel wide cutout and \texttt{supersample = 2}.\footnote{As explained in Sec. 2.1.1 of G22, the image grid can be supersampled by some integer factor, which typically is set to be 1 or 2.}

\newcommand{\scale}[2]{\textcolor{black}{#1} \ / \ \textcolor{black}{#2}}
We use the same simulation and modeling prior distributions as G22, reproduced below in Equation~\ref{eqn:sim-dist}. 
As in G22, we use a ``simulation distribution" to generate a full set of 22 parameters for 100 lensing systems, 
and a wider modeling prior to model these lenses.
\begin{equation}
\begin{split}\label{eqn:sim-dist}
\text{Lens mass}: &
\left\lbrace
\begin{array}{@{}r@{\quad}l@{}r@{}}\theta_{E} &\sim \exp(\mathcal{N}(\ln 1.25, \scale{0.25}{0.4})) & \hspace{78 pt} \text{[Einstein radius(\twopr)]} \\
\gamma_{epl} &\sim \mathcal{TN}(2, \scale{0.25}{0.5}; 1, 3) & \text{[Mass slope]} \\
\epsilon_{epl,1}, \epsilon_{epl,2} &\sim \mathcal{N}(0, \scale{0.1}{0.2}) & \text{\qquad [Lens mass eccentricities]} \\
x_{epl}, y_{epl} &\sim \mathcal{N}(0, \scale{0.03}{0.06}) & \text{[Lens mass center(\twopr)]} \\
\gamma_{ext,1}, \gamma_{ext,2} &\sim \mathcal{N}(0, \scale{0.05}{0.1}) & \text{[External shear components]}
\end{array} \right. \\
\text{Lens light}: &\left\lbrace
\begin{array}{@{}r@{\quad}l@{}r@{}}R_{l} &\sim \exp(\mathcal{N}(\ln 1.6, \scale{0.15}{0.25})) & \hspace{65 pt}  \text{[Lens S\'ersic radius(\twopr)]} \\
n_{l} &\sim \mathcal{U}(\scale{2}{0.5}, \scale{6}{8}) & \text{[Lens S\'ersic index]} \\
\thinspace \epsilon_{l,1}, \epsilon_{l,2} &\sim \mathcal{TN}(0, \scale{0.05}{0.1}; -0.15, 0.15) & \text{[Lens light eccentricities]}\\
x_{l}, y_{l} &\sim \mathcal{N}(0, \scale{0.01}{0.02}) & \text{[Lens light center(\twopr)]} \\
I_l &\sim \exp(\mathcal{N}(\ln 300, \scale{0.3}{0.5})) & \text{[Lens half light intensity]}
\end{array} \right. \\
\text{Source light}: &\left\lbrace
\begin{array}{@{}r@{\quad}l@{}r@{}}R_s &\sim \exp(\mathcal{N}(\ln 0.25, \scale{0.15}{0.25})) & \hspace{65 pt} \text{[Source S\'ersic radius(\twopr)]} \\
n_s &\sim \mathcal{U}(0.5, \scale{4}{8}) & \text{[Source S\'ersic index]} \\
\thinspace \epsilon_{s,1}, \epsilon_{s,2} &\sim \mathcal{TN}(0, \scale{0.15}{0.3}; -0.5, 0.5) & \text{[Source light eccentricities]}\\
x_s, y_s &\sim \mathcal{N}(0, \scale{0.25}{0.5}) & \text{[Source light center(\twopr)]} \\
I_s &\sim \exp(\mathcal{N}(\ln 150, \scale{0.5}{0.9})) & \text{[Source half light intensity]}
\end{array} \right.
\end{split}
\end{equation}

\noindent
where $\mathcal{U}(a,b)$ is a uniform distribution with support $\qty[a,b]$, $\mathcal{N}(\mu, \sigma)$ is Gaussian with mean $\mu$ and standard deviation $\sigma$, and $\mathcal{TN}(\mu, \sigma; x_{low}, x_{high})$ is a truncated Gaussian with support $\qty[x_{low}, x_{high}]$. For the distribution parameters, we use the notation $\scale{a}{b}$ to indicate that the simulation distribution uses the parameter $\textcolor{black}{a}$ while the prior uses $\textcolor{black}{b}$. For instance, when generating our dataset, we sample $x_{epl}$ from $\mathcal{N}(0, 0.03)$, while during modeling, our prior for $x_{epl}$ is $\mathcal{N}(0, 0.06)$.


We initially ran the pipeline with the hyperparameters in Table \ref{tab:100sysparams} on all systems.

\begin{deluxetable}{ ccc }[H]
\tablecaption{Pipeline Hyperparameters for 100 Simulated Test Systems} \label{tab:100sysparams} 
\tablehead{
    MAP & SVI & HMC
}
\startdata
\hline
$n_\mathrm{steps}: 1000 $& $n_\mathrm{steps}: 5000$ & $n_\mathrm{sample}: 1500$ \\ 
$n_\mathrm{MAP}: 2000$ & $n_\mathrm{VI} : 1000$ &  $n_\mathrm{burn} : 500$ \\ 
&  & $n_\mathrm{HMC}: 64$ \\ 
\enddata
\end{deluxetable}

\begin{figure}[H]
    \centering
    \includegraphics[width=1.0\linewidth]{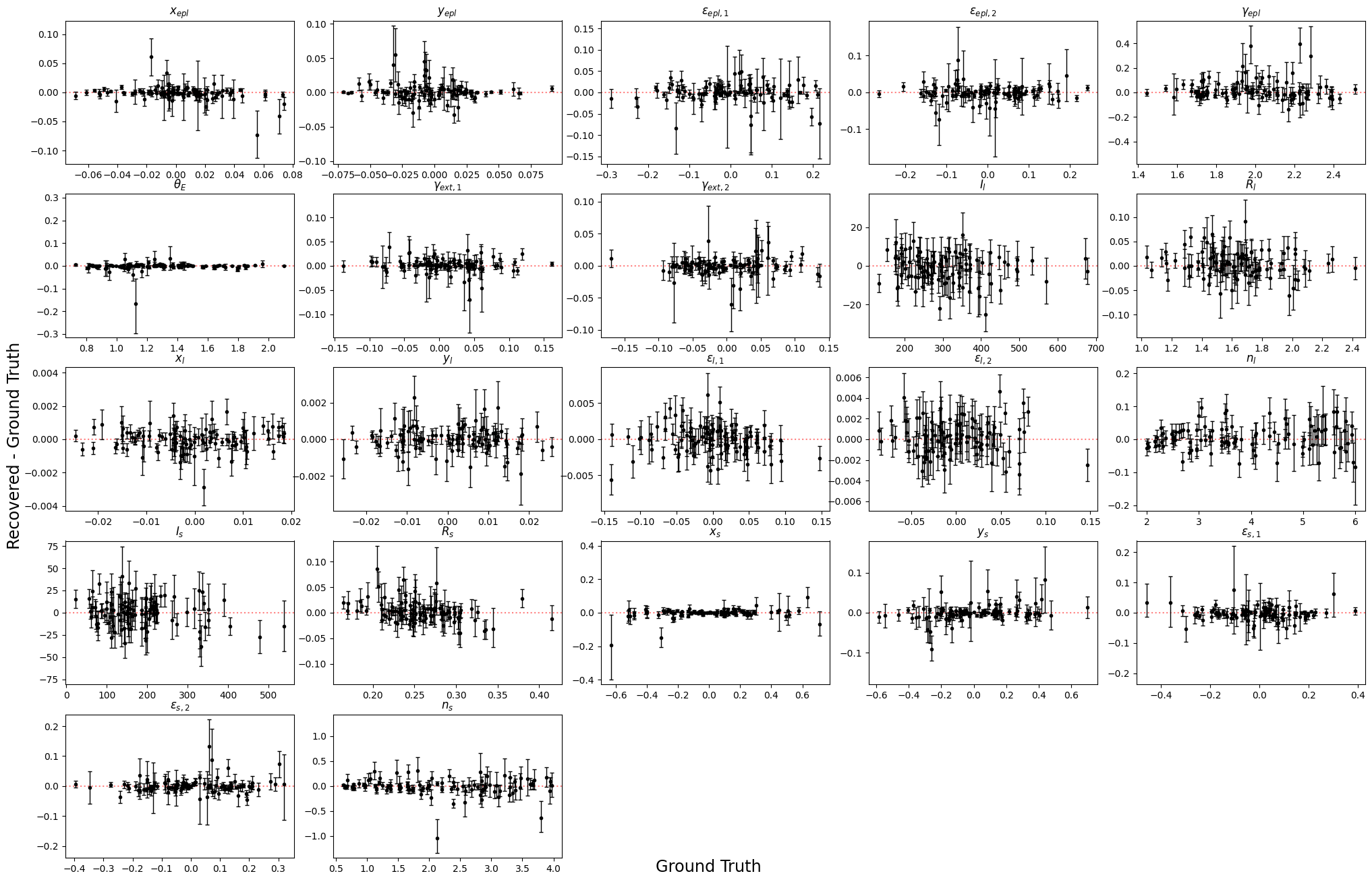}
    \caption{Modeling residuals for 100 simulated systems, with the true parameter values on the $x$-axis and residuals on the $y$-axis. Error bars represent $1\sigma$ in the posterior.}
    \label{fig:100systemsresiduals}
\end{figure}

With these, a small number of systems did not converge to $\hat{R} < 1.1$. For these, we reran HMC with $n_\mathrm{sample}= 7000$ and $n_\mathrm{burn} = 2000$. 
After this process, all systems except one converged to $\hat{R}< 1.1$, and this final system (System~60) also converged after applying mass matrix adaptation (\S\,\ref{sec:mass-matrix}).
The details are provided in Appendix~\ref{sec:sys60}.
The results for all 100 systems are summarized in Figure~\ref{fig:100systemsresiduals}.

\clearpage
\section{Conclusion}\label{sec:conclusion}
All three stages of our modeling pipeline show very high paralellizability. 
In our test on the 100 simulated systems, the speed and memory gains from parallelization, along with other new techniques, allowed us to achieve convergence on even very challenging systems.

Achieving convergence using multiple GPU nodes is significantly faster than using the single node implementation. Furthermore, algorithmic improvements in \gigaltwo contribute an additional 10 – 20\%. 
As a concrete example, we demonstrate with a real system (\desitwothreeeight) that the overall execution time sees a five-fold speed gain with eight nodes versus a single node. This comparison is summarized in Table~\ref{tab:real_n-node_comp}.
Furthermore, when using \gigalone and 1 GPU node to model real lensing systems in \cite{huang2026a, huang2025c}, we used \texttt{supersample = 1}. In this work, we accomplished the modeling with \texttt{supersample = 2} for the first time  for \hst imaging data.
Finally, \cite{huang2026a, huang2025c} achieved $\hat{R} <1.1$ for all parameters, but in this work, we have achieved $\hat{R} <1.01$ for all parameters for a real system, for the first time.

The massive parallelization enabled by \gigaltwo is especially valuable for modeling systems with large cutouts and high-dimensional parameter spaces, such as the Cosmic Carousel Lens.
\citep{sheu2024a, odonnell2026a, urcelay2026a}. 
We want to emphasize that the advantage of speed gain is not primarily for the sake of speed alone, though that certainly is a benefit.
The main advantage of speed gain is to pursue rigor and accuracy. 
In the example of the Cosmic Carousel Lens, as well as  other systems, the speed gain would allow us to rigorously test systematics by iterating through different model assumptions.

\section*{Author Contributions}

X.~Huang conceived and supervised the project and, together with L.~Upson, prepared the manuscript.
E.~Liu worked on the first iteration of the implementation.
N.~Ratier-Werbin and S.~Xu wrote the original multi-node implementation.
N.~Ratier-Werbin and L.~Upson wrote the code documentation.
L.~Upson coordinated work on the paper alongside X.~Huang, and performed systematic testing.
H.~Lu wrote the latest implementation of the multi-node code.
E.~Yap managed code and environment versions, and tested the code on simulated and real systems.
E.~Odell tested the multi-node code on simulated and real systems up to 512 GPUs, and modeled \desitwothreeeight with S.~Xu.
A.~Parke tested the multi-node code, updated telemetry and output standards, and wrote the new demo notebooks.
H.~Ambardekar performed thorough code review.
N.~Demeure and B.~Richardson provided technical assistance.
A.~Gu, S.~Baltasar, and Y.~Hsu provided review and feedback during the development process.

\newpage
\section*{Acknowledgments}\label{sec:acknowledgement}
X.H. gratefully acknowledges financial support from NASA
through grant HST-GO-15867, and the University of San
Francisco Faculty Development Fund. Support for HST
program 15867 was provided by NASA through a grant from
the Space Telescope Science Institute, which is operated by the
Association of Universities for Research in Astronomy, Inc.,
under NASA contract NAS 5-26555.
This research used resources of the National Energy Research Scientific Computing Center (NERSC), a U.S. Department of Energy Office of Science User Facility operated under contract No. DE-AC02-05CH11231, and the Computational HEP program in the Department of Energy's Science Office of High Energy Physics provided resources through the ``Cosmology Data Repository'' project (grant \#KA2401022).

This material is based on work supported by the U.S. Department of Energy (DOE), Office of Science, Office of High Energy Physics, under contract No. DE-AC02-05CH11231, and by the National Energy Research Scientific Computing Center, a DOE Office of Science User Facility under the same contract. 


Any opinions, findings, and conclusions or recommendations expressed in this material are those of the authors and do not necessarily reflect the views of the U. S. Department of Energy, or any of the listed funding agencies.

\software{
    TensorFlow \citep{TensorFlow},
    TensorFlow Probability \citep{dillon2017a}, 
    JAX \citep{bradbury2018a}, 
    Optax \citep{optax2020},
    lenstronomy \citep{birrer2018a},
    Matplotlib \citep{hunter2007a},
    photutils \citep{bradley2023a}
    seaborn \citep{waskom2021a},
    corner.py \citep{foreman2016a},
    NumPy \citep{harris2020a}
}

\bibliographystyle{aasjournal}
\bibliography{dustarchive}

\appendix

\section{Simulated System 60}\label{sec:sys60}
The one system (System 60) of the 100 simulated systems in 
\S\,\ref{sec:100-sys} that requires additional refinement for the model to converge is an example that demonstrates that for some systems, a greater than typical computational power is required.

\begin{figure}[H]
    \centering
    \includegraphics[width=1\linewidth]{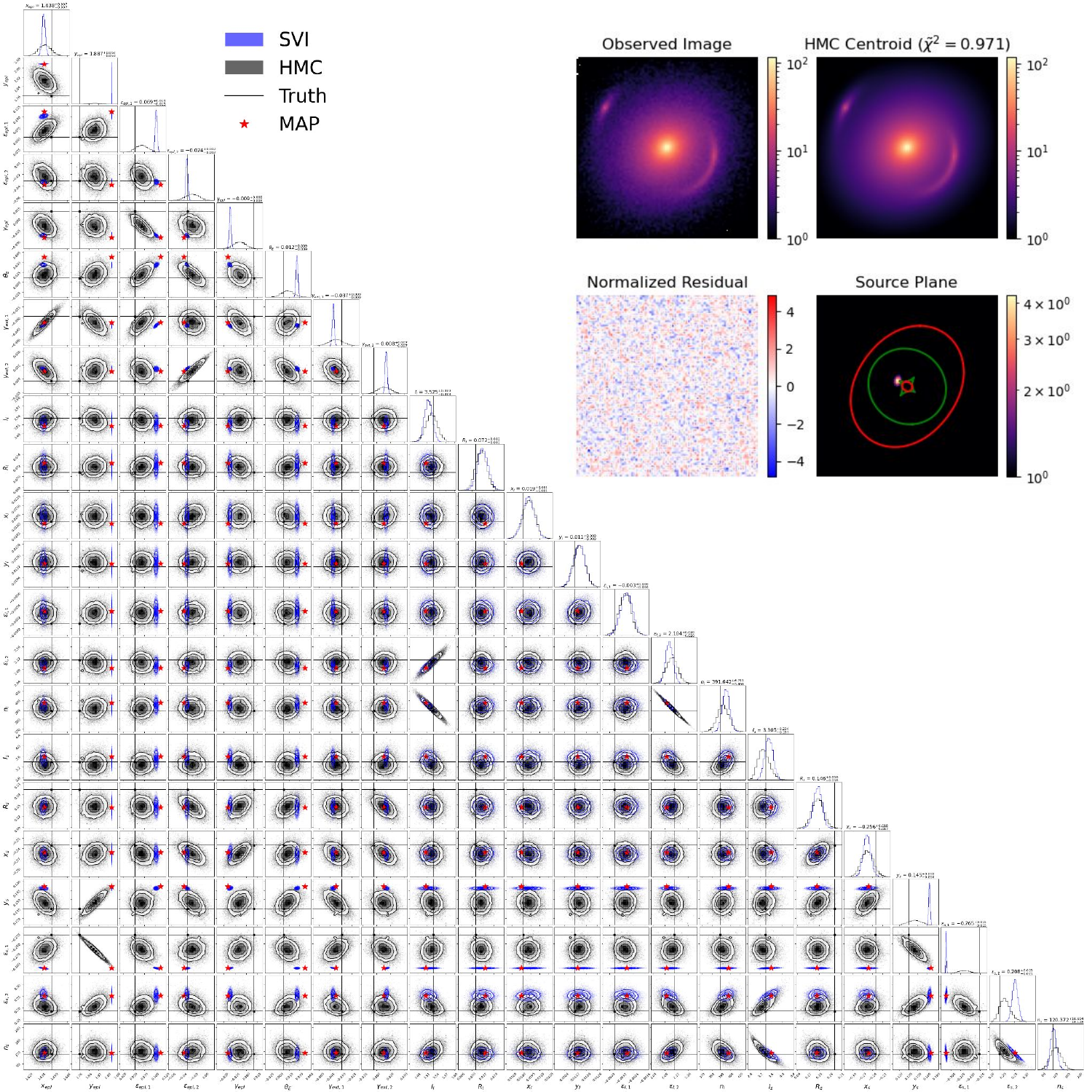}
    \caption{The modeling results for System 60. Black and blue contours are for the HMC-sampled posterior SVI surrogate, respectively. Black crosshair shows the ground truth and the red stars, best MAP results. Note that for all the mass parameters, the covariance of the SVI surrogate is significantly smaller than that of the HMC-sampled posterior, and the mean is often far the the HMC posterior's centroid and ground truth by up to $\sim 3 \, \sigma$.}
    \label{fig:sys60results}
\end{figure}

The difficulties encountered with this system stemmed from a poor surrogate from SVI. 
For this system, SVI optimized its objective correctly, finding a surrogate for a local minimum with even lower ELBO loss than a Gaussian fit to the true posterior (global minimum). However, this surrogate was very different from the true posterior (Figure~ \ref{fig:sys60results}).\footnote{In \citet{huang2026a}, we already noticed that for the real systems, DESI~J165.4754-06.0423, there is an offset between the SVI surrogate and HMC posterior, indicative of a poor surrogate. However, the discrepancy was not as severe as in this simulated system.} 
Furthermore, we found the same SVI behavior even if we start SVI from the ground truth.
With this incorrect surrogate, HMC struggled to converge under our ``standard'' pipeline setting of $n_{\mathrm{sample}}=1500$, $n_{\mathrm{burn}}=500$.
However, we are able to achieve HMC sampling convergence to the true posterior with $\hat{R}<1.1$ by using mass matrix adaptation (\S\,\ref{sec:mass-matrix}),
with $n_\mathrm{mass\_mat\_adapt} = 40000$ samples, 
and a large number of final sample steps 
($n_\mathrm{sample}= 250000$).

The performance improvement we have gained from the multi-node implementation presented in this work is critical in achieving convergence for this system. 


\section{Demonstration of \gigaltwo on a Real System}\label{sec:real-sys}

Below, we do a direct comparison between modeling results and compute time on a single node versus eight nodes. \desitwothreeeight was first modeled on a single GPU node using the original \gigal in \citet[][H25]{huang2025c}. 
For this comparison, we redo the single-node modeling using \gigaltwo rather than using the compute time reported in H25. This is because the single-node execution time in the present work is substantially faster than in H25, due to improvements in the inference algorithm, differences in optimizer choice, and other implementation updates; this speedup is achieved despite the use of supersampling in the present analysis, which was not used in H25.


On a single node, the whole modeling process took more than two hours. On eight nodes, the modeling process took just over 25 minutes, a five-fold compute time speedup, with a majority of the improvement from HMC (Table \ref{tab:real_n-node_comp}). The best-fit parameters for each model agree within a fraction of the parameter uncertainties, and the resulting reduced $\chi^2$ is the same to 4 decimal places at $\chi^2_\nu=0.8954$.

Furthermore, we achieved convergence with the backward model with $\hat R<1.01$ in all 38 parameters, compared to H25 achieving only $\hat R<1.1$ in all parameters. To achieve convergence with $\hat R<1.01$, we used a wider prior than H25 . The widened parameters were $R_{ext}$, $n_{ext}$, $n_{S1}$, and $n_{S2}$. In addition, for the first time we used \texttt{supersample = 2} for a real system with \hst data. By comparison, H25 used \texttt{supersample = 1} for this system and for the other five systems with \hst data. 

\begin{deluxetable}{lcccc}[H]
\tablecaption{Real System Single-Node Performance vs Eight Node Performance in \gigaltwo} \label{tab:real_n-node_comp}
\tabletypesize{\footnotesize}
\renewcommand{\arraystretch}{1.4}
\tablehead{
    \colhead {Nodes} &
    \colhead{Total time} &
    \colhead{Supersample} &
    \colhead{$\hat{R}$ for all parameters}
}
\startdata
1 & 128~min and 24~sec & 2 & $<1.01$ \\
8 & 25~min and 12~sec & 2 & $<1.01$ 
\enddata
\end{deluxetable}

\begin{figure}[H]
    \centering
    \includegraphics[width=0.9\linewidth]{figs/cornerplot-desi238-pngs.pdf}
    \caption{Corner plot of the eight-node \gigaltwo model of \desitwothreeeight. We show the full corner plot for all 38 parameters, simultaneously modeled and sampled. For presentation purposes, every 10th sample is plotted.
    In order, the parameters shown are: the mass parameters for the lens, the \ser profile for a field (``environmental'') galaxy with subscript ``env'', two \ser profiles for the main lens with subscripts of ``L1'' and ``L2'' for the two components of its light profile, and \ser profile parameters for each of the two lensed sources with subscripts ``S1'' and ``S2''. 
    For visual clarity, in the inset, we show again the corner plot of just the mass parameters.}
    \label{fig:238corner}
\end{figure}
\begin{figure}[H]
    \centering
    \includegraphics[width=1\linewidth]{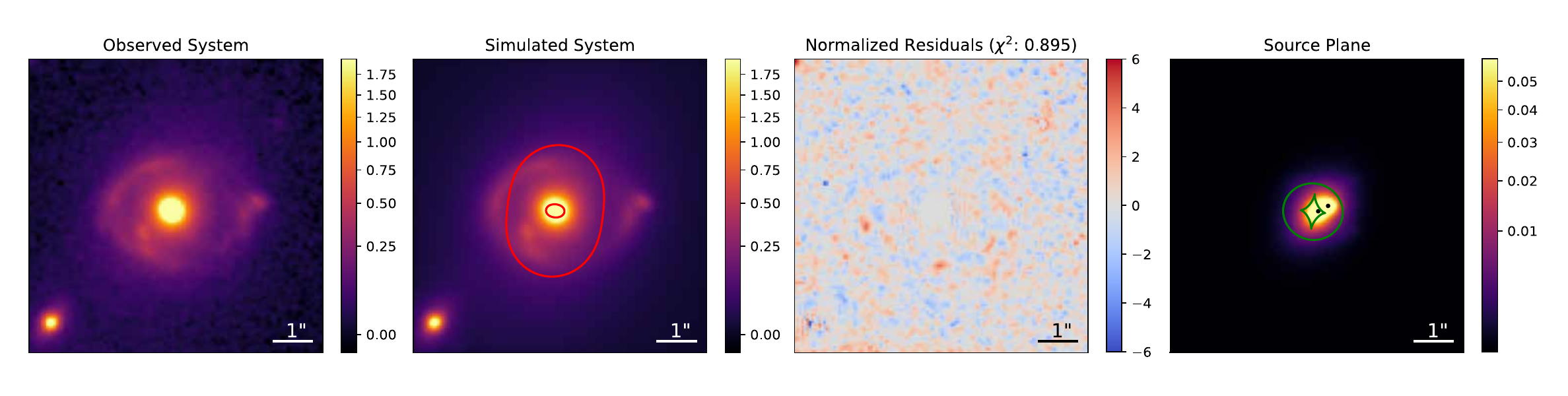}
    \caption{Best-fit model for the eight-node model of \desitwothreeeight. From left to right: The observed image from \hst, the best-fit model with the critical curves, the normalized residual map, and the forward-modeled unlensed sources, with the centers of the two sources marked by black dots and the caustics overlaid.}
    \label{fig:238model}
\end{figure}

For this real system with 38 free parameters, we have reached $\rhat < 1.01$.
Reaching $\rhat < 1.01$ typically requires substantially longer runtimes than to reach $\rhat < 1.1$, by roughly an order of magnitude even for systems with far fewer parameters \citep{baltasar2026}.
But with \gigaltwo using multiple GPU nodes we  demonstrate that such stringent convergence is feasible for strong lensing, and with continued improvements in software and hardware, may become increasingly attainable for a broader range of strong-lensing applications.


\section{Implementation Details}\label{sec:app-implement-details}
\subsection{JAX's \shardmap API}

For distributed computation across multiple GPU nodes, we use \shardmap from JAX.\footnote{\url{https://docs.jax.dev/en/latest/jep/14273-shard-map.html}} 
Below we provide a brief description of \shardmap. 
The behavior of \shardmap is determined by \inspec and \outspec. First, a function needs to be defined for a block of data. \shardmap then shards the data  into chunks as specified by \inspec, and passes it into the function. The function performs computation on each chunk and returns the result. \outspec specifies how the output is to be to reassembled  from each shard into a final array.

If \outspec has the same sharded dimensions as \inspec, it just reassembles the array the same way it was split up for \inspec. If it has less, then the result is assumed to be replicated across the axes that were sharded in \inspec but are not in \outspec---e.g., because there was a summing (\texttt{psum}) or averaging 
(\texttt{pmean}) operation carried out along one of these axes---and only one of the final copies across that axis is returned from \shardmap. If \outspec has more sharded axes than \inspec, \shardmap will effectively tile the result of the function evaluation across the axes that were not sharded in the input but are in the output.
Communication between shards is facilitated by the use of special functions that gather or aggregate arrays across shards (e.g., \texttt{pmean}, \texttt{psum}).

\subsection{\shardmap Not Used for HMC}

HMC, as implemented in TensorFlow Probability's JAX backend, assumes that scalar quantities such as step sizes and adaptation statistics have certain scalar types on each device. 
In JAX, this corresponds to values without explicit sharding annotations. By contrast, \shardmap enforces a model in which the data type contains explicit sharding labels---information about how the array is sharded.\footnote{https://docs.jax.dev/en/latest/notebooks/explicit-sharding.html} This mismatch leads to data type (\texttt{dtype}) and array shape inconsistencies, and arises from JAX's type and sharding semantics. This cannot be resolved without reimplementing HMC to be sharding label-aware. We therefore use \texttt{pmap} for per-device replicated execution and \texttt{jax.experimental.multihost\_utils.process\_allgather} to aggregate samples across hosts.

\section{Containerized Execution with Docker and Shifter}\label{sec:app-docker}
Initially, we encountered issues where different nodes or Jupyter sessions on Perlmutter would load slightly different CUDA, cuDNN, or NCCL\footnote{NCCL is a library developed by NVIDIA for GPU-GPU communication in high-performance computing; see \citet{hu2025} for more details.} builds, leading to failures because certain packages were not compatible with one another. Instead of attempting to manually reconcile these dependencies, we pulled an official, version-locked CUDA base image and built our entire software stack on top of it.  To ensure that this environment is used consistently across all workflows, we then created a custom Jupyter kernel that launches entirely within our \gigaltwo multi-node Docker/Shifter image. 
To be able to use our container\footnote{We do this by using Shifter, \url{https://shifter.readthedocs.io/en/latest/}.} with GPU and NCCL support, we modify the configuration for Jupyter appropriately.\footnote{This is done through specifying \texttt{kernelspec}, which defines the behavior of the backend process (called a kernel) responsible for running code in a Jupyter notebook. For more details, see \url{https://jupyter-client.readthedocs.io/en/stable/kernels.html\#kernelspecs}.} Using this container, both interactive Jupyter notebooks and batch jobs\footnote{For details, see \url{https://docs.nersc.gov/jobs/}.} can run inside the exact same controlled environment. This approach eliminates dependency drift and guarantees fully reproducible, multi-node execution on NERSC.


\end{document}